# Oxygen Functionalization-induced Crossover in the Tensile Properties of thinnest 2D Ti$_2$C MXene

Hongzhi Yu[1], Ke Xu[2], Zhisen Zhang[2], Jian Weng[1], Jianyang Wu[2,*]

[1]Department of Biomaterials, College of Materials, Xiamen University, Xiamen 361005, China

[2]Department of Physics, Research Institute for Biomimetics and Soft Matter and Jiujiang Research Institute, Xiamen University, Xiamen 361005, PR China

**Abstract:** Transition metal carbides/nitrides (MXenes) are a newly developing class of two-dimensional (2D) materials with technically robust properties that can be finely tuned by planar surface functionalization. Herein, the critical role of oxygen (O-) functionalization on the tensile mechanical characteristics of thinnest 2D Ti$_2$C MXene is explored by molecular dynamic (MD) simulation with first-principle based ReaxFF forcefield. It is demonstrated that Ti$_2$C sheet shows unique tensile mechanical behaviors that pronouncedly vary with the content of O-functionalization and stretching direction. Upon both loading directions, there is an apparent crossover in the Young's modulus, failure strength and failure strain. Intriguingly, under armchair directional load, a structural transition of 1T to 1T' phase occurs in the Ti$_2$C region, which has been observed in many transition metal dichalcogenides. Upon zigzag directional straining, however, two distinct structural transformations take place in pristine and fully O-functionalized Ti$_2$C sheets, respectively. As the load is removed, those three structural transformations are reversible, and they are critically understood by analysis of the bond configurations. The study provides important insights into mechanical behaviors and structural transformations of functionalized MXenes.

## 1. Introduction

[*]Corresponding Email: jianyang@xmu.edu.cn

Beyond graphene, phosphorene and transition metal dichalcogenides (TMDs), another family of two-dimensional (2D) freestanding materials called transition metal carbides/nitrides (MXenes) was discovered in 2011[1]. MXenes were commonly produced by selectively wet-chemical etching from the layered parent ternary carbide and nitrides crystal, named MAX phases[1, 2]. The MAX phases are anisotropic laminated structure with a general formula of $M_{n+1}AX_n$, where $n$ = 1, 2 or 3, M, A and X represent transition metals, a group of 13 or 14 element, and carbon and/or nitrogen, respectively. Because of a large number of combinations of the three elements in the MAX phase, over 70 kinds of MXenes have been experimentally synthesized/theoretically predicted[3].

Due to their unique structural, physical and chemical properties, MXenes found a myriad of important practical applications, such as sensors[4, 5], energy storage [6], catalysts[7], electromagnetic shielding[8], and so on[2]. For example, $Ti_3C_2T_x$ MXene film acts as metallic channels for chemiresistive gas sensors with ultrahigh sensitivity[4], while single-layer $Ti_3C_2$ MXene can be used to construct the $NH_3$ sensor with high selectivity[5]. As a result of the fact that metallic ions can be inserted into layered MXenes, MXene nanosheets serve as a promising anode material that can be recharged with a variety of cations, such as $Li^+$, $Na^+$, $K^+$, $Ca^{2+}$, $Mg^{2+}$ and $Al^{3+}$[9-14]. PtCo nanoparticles (NPs) supported-layered $Ti_3C_2X_2$ has been successfully adopted as a catalyst for the hydrolysis of Ammonia borane (AB) [7]. It was reported that a variety of MXenes including monometallic MXenes, ordered double-metallic MXenes, and random solid solution MXenes are able to achieve effective electromagnetic interference (EMI) shielding in micrometer-thick films[8].

Because MXenes show great potential for a variety of applications, fundamental mechanical properties have been of major interest to material scientists and engineers. Using atomic force microscopy (AFM) measurements, it was reported that elastic modulus of $Ti_3C_2T_x$ is 330 ± 30 GPa[15].

Using first-principles calculations, mechanical stress-strain curves and deformation mechanisms of 2D $Ti_{n+1}C_n$ in response to tensile loads, as well as the rigidity of 2D MXenes with different functional groups, were examined[16, 17]. Young's modulus and in-plane stiffness of 2D $Ti_2CO_2$ and $Ti_3C_2O_2$ are reduced with the increase of temperature[18]. Moreover, through first-principles calculations, elastic stiffnesses of 2D $Ti_3C_2/Ti_2C$ greatly depend on the terminated functional groups; both functional group-free and oxygen (O)-terminated MXenes exhibit high elastic stiffness, while MXenes with other functional groups show low elastic stiffness[19]. Using classic molecular dynamics (MD) simulations, the influence of point defects on the nanoindentation mechanical responses of $Ti_{n+1}C_nT_x$ MXenes were investigated[20]. Using MD simulations with an optimized forcefield, it was shown that the thinnest $Ti_2C$ yields the highest Young's modulus due to the folding and crimp caused by the high surface energy, and the failure and rupture patterns depend on the number of atomic layers[21]. Moreover, bending deformation of $Ti_{n+1}C_n$ MXenes via classical MD calculations showed that $Ti_2C$ is more resistant towards bending than graphene[22].

O-functionalized $Ti_{n+1}C_nT_x$ show unique properties compared to other surface terminated ones [23-27]. For example, O-functionalized $Ti_3C_2T_x$ and $Ti_2CT_x$ exhibit the highest theoretical $Li^+$ storage capacities[23]. $Ti_2CO_2$ shows semiconductor character, while $Ti_2CF_2$ and $Ti_2C(OH)_2$ exhibit metallic character[24]. In addition, O-functionalized $Ti_2CT_x$ shows higher thermodynamic stability than -OH and -F groups-surface functionalized ones[25-27]. Despite great efforts dedicated to the thinnest 2D $Ti_2C$ MXene, its mechanical properties, particularly the tensile behavior and the effect of O-functionalization, are remain largely unknown yet. To this end, the objective of this study is to reveal the critical role of O-surface termination on the tensile mechanical characteristics of thinnest 2D $Ti_2C$ MXene using MD simulations with first-principle based ReaxFF forcefield.

## 2. Models and Methodology

### 2.1 Atomic Structures

Ti$_2$C is the first synthesized thinnest MXene. The initial atomic model of pristine Ti$_2$C MXene is generated by deleting aluminum (Al) layers from the parent bulk Ti$_2$AlC MAX phase. The titanium (Ti) and carbon (C) atoms of Ti$_2$C are placed in the lattice positions as 4$f$ (1/3, 2/3, $u$) and 2$a$ (0, 0, 0), respectively[28]. Such 2D MXene material is structurally characterized by three atomic layers with a hexagonal-like unit cell, in which the carbon layer is sandwiched between two Ti monoatomic hexagonal-planes, as shown in Figure 1a and b. The Ti$_2$C MXene monolayer with oxygen functional group is constructed with the most energetically-stable configuration[18], as shown in Figure 1c and d. Note that the functionalization of O-functional group on the surfaces of Ti$_2$C varies from 0-100%. For a given content of O-functionalization, ten samples with random distributions of O-functionalization are created for statistically capturing the tensile mechanical characteristics. To investigate the tensile mechanical properties of O-functionalized Ti$_2$C MXene monolayers, their 2D lateral dimensions are created to be around 60.0 × 60.0 Å$^2$.

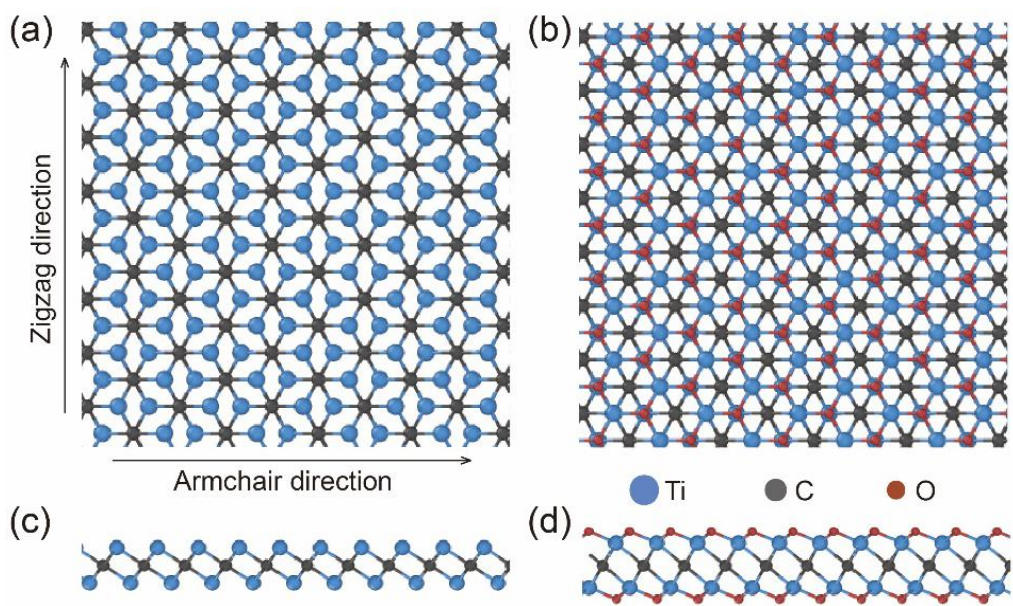

Figure 1 Atomic models of thinnest 2D Ti$_2$C-based MXenes. (a) and (b) Top-views of Ti$_2$C and Ti$_2$CO$_2$ sheets, respectively. (c) and (d) Side-views of Ti$_2$C and Ti$_2$CO$_2$ sheets, respectively.

**2.2 Tension MD simulations**

Prior to the uniaxial tension, all free-standing Ti$_2$CO$_x$ sheets are initially quasi-statically relaxed to a local minimum-energy configuration by conjugate gradient method with an energy tolerance of $1.0 \times 10^{-4}$ kcal/mole and a force tolerance of $1.0 \times 10^{-4}$ kcal/(mole·Å), respectively. Then, MD relaxations of 100000 timesteps are performed at zero pressure and temperature of 10 K along planar directions under *NPT* ensemble (constant number of particles, constant pressure, and constant temperature) for achieving stable samples. Initial velocities of Ti and C atoms in the MXene systems are assigned according to Gaussian distribution with the given temperature. Note that temperature of 10 K is chosen to avoid the effects of thermal fluctuations. Finally, uniaxial tension along one of planar directions is implemented by deformation control method under an *NPT* ensemble, and a reasonable strain rate of $1.0 \times 10^8$/s is utilized. The strain increment is imposed to as-relaxed samples every 1000 timesteps. Such simulation settings take into consideration of Poisson effect that allows expansion/contraction in the loading-free planar direction. In those MD simulations, Nosé-Hoover thermostat and barostat methods with a damping time of 100 and 1000 timesteps are used to control the temperature and pressure, respectively. A small timestep of 0.1 fs with the velocity-Verlet integration method is applied, ensuring stable MD simulations. Periodic boundary conditions (PBCs) are imposed along the two planar directions to mimic an infinite sheet, which is able to eliminate any spurious effects of boundaries, while non-PBCs is applied along the out-of-plane direction. Atomic stresses in the MXene systems are computed on the basis of the virial definition of stress. Atomic potential energy and atomic stress in the MXene systems are averaged every 1000 timesteps to eliminate the thermal oscillations during the MD simulations.

## 2.3 Forcefield for $Ti_2CO_x$ MXene Systems

All the classic MD simulations are implemented using the Large-scale Atomic/Molecular Massively Parallel Simulator (LAMMPS) package code. Previously, to describe the atomic interactions in the $Ti_2CO_x$ systems, first-principle based ReaxFF potential has been applied to investigate a large variety of systems, which is able to capture phenomena involving high-temperature thermal reactivity, the breakage and re-formation events of covalent bonds during mechanical loads[29-32]. Moreover, it was showed that the ReaxFF potential is able to predict the nanoindentation and friction mechanical responses of Ti-C systems, as well as their fracture behaviors[20, 34, 35]. In this study, the version of ReaxFF forcefield developed by Osti et al.[33] is therefore adopted to simulate the tensile mechanical characteristics of 2D $Ti_2CO_x$ systems.

## 3. Results and Discussion

### 3.1 Uniaxial Tensile Mechanical Responses

Figure 2a shows the resulting uniaxial tensile stress-strain responses of pristine and O-functionalized $Ti_2C$ sheets under armchair directional load. Apparently, all samples show unique tensile mechanical responses that greatly vary with the content of surface-functionalization of oxygen atoms. Three or four deformation stages can be roughly identified from the global mechanical curves, depending on the content of O-functionalization. As the content of O-functionalization varies from 0-70%, it is identified four deformation stages, however, there are three deformation stages identified for $Ti_2C$ sheets with O-functionalization of 80-100%. The first deformation stage of all samples is characterized by the initial linear tensile responses within limit strain regimes, representing the linear elastic behaviors. The second deformation stage of all samples is described by strain-softened tensile stress-strain responses within finite strain regime, namely, reduction in the slopes of stress-strain curves with finite strains. In those strain regimes, the loading tensile stresses nonlinearly rise up to the first peaks with increase of strain in the curves, representing that they are nonlinear-elastically stretched. Remarkably, the character of nonlinear variations in the tensile stresses with elastic strain is highly related to the fraction of O-functionalization. The deformational stage III of $Ti_2C$ sheets with O-functionalization varying from 80-100% is characterized by sudden deep drops of tensile stresses from the highest peaks in the global curves, followed by a series of sudden rise-and-drop events of loading stresses, indicating occurrence of ductile failure and significant plastic deformation. In sharp contrast, the deformation stage III of $Ti_2C$ sheets with O-functionalization varying from 0-70% is described by that, prior to the highest peak of tensile stress, the loading tensile stresses non-smoothly changes with increase of strain, namely, a number of sudden drop-and-rise events of tensile stress occur. As an example, for pristine $Ti_2C$ sheet, it is interestingly observed a plateau stage between the first and highest peaks in the curve, indicating large-scale structural transformation events within this strain regime, followed by a long-range smooth rise in loading stress up to the maximum tensile strength. Similarly, the final deformational stage IV of $Ti_2C$ sheets with O-functionalization from 0-70% is characterized by sudden deep drops of stretching stresses in the curves, followed by decay in the loading stresses.

Figure 2b presents the simulated uniaxial tensile stress-strain responses of pristine and O-functionalized $Ti_2C$ sheets under zigzag directional load. Differing from the case under armchair directional load, four deformational stages can be mainly classified from the global nonlinear curves, which does not depend on the content of O-functionalization. Similar to the case of $Ti_2C$ sheets with O-functionalization of 0-70%, the deformational stages I, II and IV are dominated by the initial linear tensile responses, nonlinear tensile responses and sudden drops of tensile loading stresses in the curves for all samples, respectively. By comparison, however, there exists apparent difference in the character of nonlinearity in the curves that represent the third deformational stage. For example, for fully O-functionalized Ti$_2$C sheet under zigzag directional load, it is detected a short plateau phase in the deformational stage III, whereas for the case under armchair directional straining, the deformational stage III is dominated by sudden deep drop of loading stress. Upon zigzag directional stretching, as the Ti$_2$C sheet is functionalization-free, there is a clear crossover from strain-softening to strain-hardening behavior prior to sudden deep drop of loading stress, differing from the occurrence of a long plateau phase under armchair directional load. Moreover, it is a drop of loading stress to zero in the deformational stage IV, suggesting that functionalization-free Ti$_2$C sheet catastrophically fails via brittle fracture. It is summarized that there is anisotropy in the in-plane tensile mechanical responses of Ti$_2$C sheet that are tuned by the content of O-functionalization.

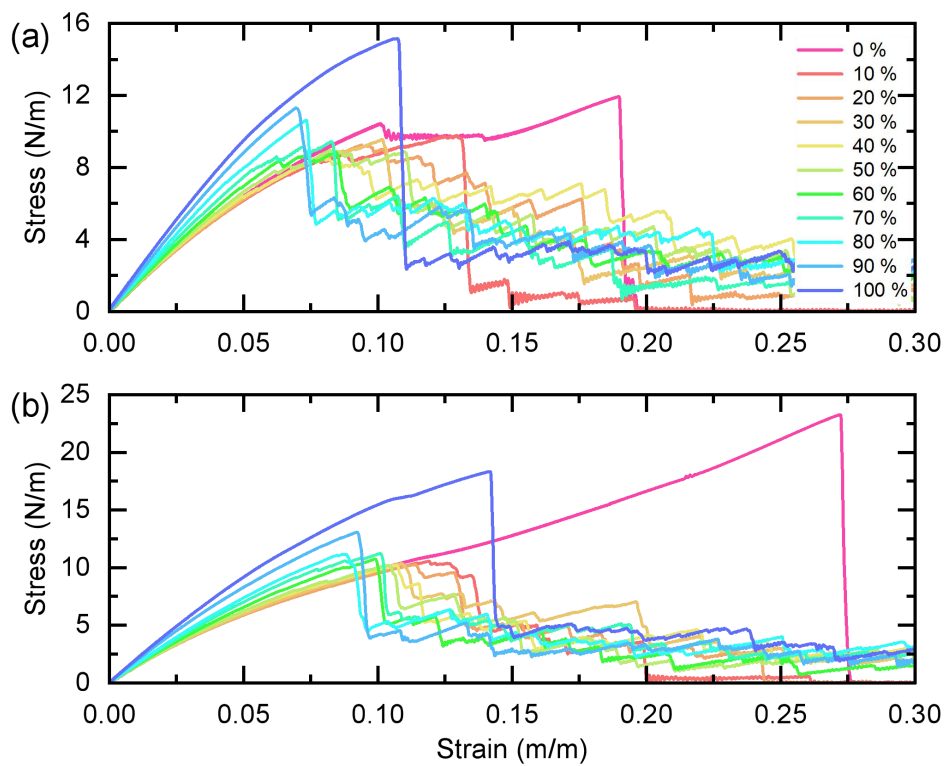

Figure 2 Uniaxial tensile stress-strain response of Ti$_2$C sheets functionalized by oxygen atoms on the surfaces varying from 0-100 % subjected to (a) armchair and (b) zigzag directional loads, respectively.

## 3.2 Oxygen-functionalization induced Crossover in the Mechanical Properties

Fundamental tensile mechanical properties of 2D $Ti_2CO_x$ systems can be extracted from the mechanical loading curves shown in Figure 2. Figure 3a plots the predicted Young's modulus as a function of the content of O-functionalization, respectively. Clearly, upon on both loading directions, the Young's modulus strongly depends on the content of O-functionalization. Intriguingly, subjected to both planar loading directions, 2D $Ti_2CO_{0.6}$ sheet is the most mechanically soft structure, namely, there is a crossover in the Young's modulus as the content of O-functionalization is 30%. As the content of O-functionalization is below 30%, upon armchair directional load, there is negligible change in the Young's modulus, whereas under zigzag directional stretching, the Young's modulus reduces with increasing the content of O-functionalization. As the content of O-functionalization is over 30%, upon both directional loads, the Young's modulus nonlinearly increases with increasing the content of O-functionalization, and there is negligible difference in the Young's moduli between both directional loads. As is known, the standard molar formation enthalpy of Ti-O crystalline system is -518.4 kJ/mole, whereas for Ti-C crystalline system, it is -186.2 kJ/mole. Such big difference in the formation enthalpies between Ti-O and Ti-C systems is mainly responsible for the O-functionalization tuned Young's modulus of TiC sheet. From molecular structural point of view, few oxygen atoms asymmetrically functionalized on planar surfaces locally distort the parent Ti-C bond configurations of $Ti_2C$ sheet, thereby mechanically softening internal Ti-C bond configurations; however, a large fraction of oxygen atoms chemically covered on the planar surfaces enhance the Young's modulus as a result of relative well-distributed stiffer Ti-O bonds. For example, as the $Ti_2C$ sheet is ideally O-functionalized, the applied elastic strains are dominated by the symmetrical Ti-O bonds on the planar surfaces, resulting in the stiffest responses to both armchair and zigzag directional loads.

Figures 3b and c show variations in the failure strength and failure strain with the content of O-functionalization on the Ti$_2$C sheet, respectively. Similarly, both failure strength and failure strain are finely tuned by O-functionalization. It is revealed that, upon both armchair and zigzag directional loads, O-functionalization on the planar surfaces either mechanically weakens or enhances the Ti$_2$C sheet, as well as enhances or degrades the failure strain, depending on the content of O-functionalization. As the Ti$_2$C sheet is half-functionalized by oxygen, there is a crossover in both failure strength and failure strain. As is known, the arrangement of oxygen atoms on the planar surfaces determines the symmetry of O-functionalized Ti$_2$C sheet. In this case, the Ti$_2$C sheet with random O-functionalization of 50% has the biggest possibility to show the worst regularity and uniformity in the atomic arrangement of oxygen on the planar surfaces, explaining the crossover in both failure strength and failure strains. On the contrary, both the pristine and perfect-functionalized Ti$_2$C sheets show good symmetry in atomic arrangement, resulting in uniform distribution of loading tensile stress and thereby significantly enhancing failure strength and failure strain. In addition, fully-O-functionalized Ti$_2$C sheet yields a higher failure strength than the pristine Ti$_2$C sheet due to its stronger Ti-O bonds on the planar surfaces. Interestingly, as a result of its 1T crystalline phase, Ti$_2$CO$_x$ sheets are more mechanically robust under zigzag directional load than under armchair directional load, which is opposite to the case of 2H crystalline graphene.

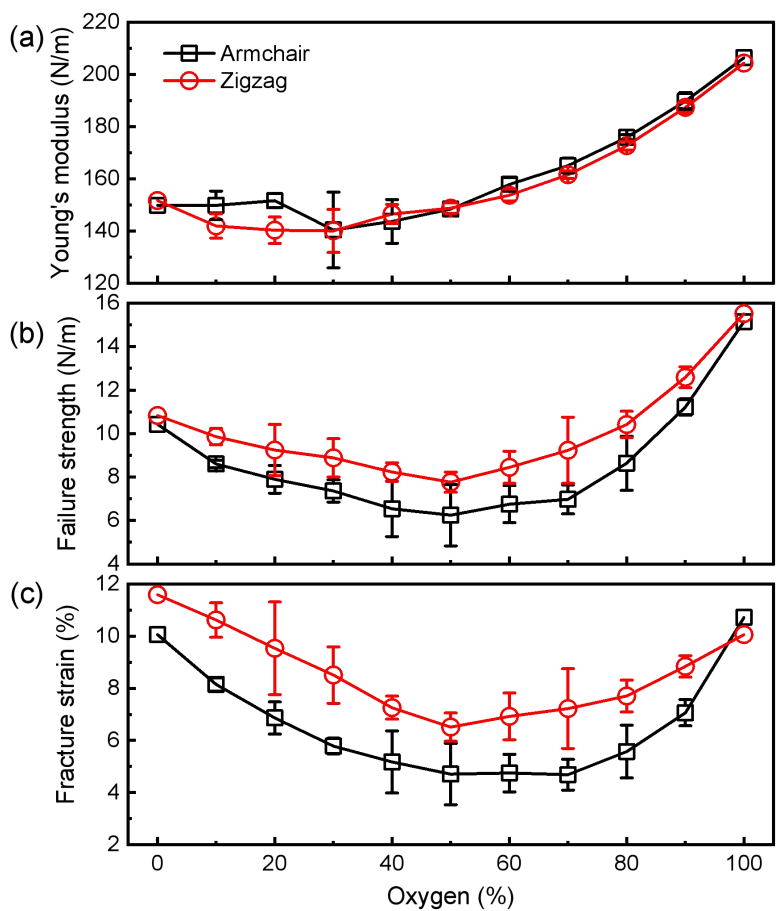

Figure 3 Uniaxial tensile mechanical properties of Ti$_2$CO$_x$ sheets. (a)-(c) Variations in the Young's modulus, failure strength and failure strain with the content of oxygen-functionalization on the surfaces for thinnest 2D Ti$_2$C MXene under both armchair and zigzag directional loads, respectively.

### 3.3 Structural Transformations and Deformation Mechanisms

To provide atomistic insights into the deformation mechanisms underlying O-functionalization controlled mechanical behaviors of Ti$_2$C sheets, structural evolution of Ti$_2$CO$_x$ sheets subjected to both armchair and zigzag directional tensile loads are captured. Figures 4a-d present the representative top-viewed snapshots of specific Ti$_2$CO$_x$ sheets ($x$ = 0.0 and 2.0) under both armchair and zigzag directional elongations at different strains, respectively, in which the insets highlight the local atomic structures and atoms are rendered on the basis of values of shear strains. With regard to functionalization-free Ti$_2$C sheet, as the armchair directional strain reaches to critical value (around

0.100), it is intriguingly identified an occurrence of phase transition from 1T to 1T' phase. Such phase transition proceeds step-by-step and completely accomplishes on the entire area of sheet as the strain is applied to around 0.150. As a result, the unit cell of Ti$_2$C sheet elongates from 5.3 to 6.1 Å, as shown in the insets of Figure 4a. Such strain-induced phase transition has been not reported in previous studies[35]. However, in 2D TMDs, it has been experimentally/numerically identified phase transition from 1T to 1T' structure under a variety of conditions such as high temperature and mechanical straining[36-39]. This step-by-step phase transition explains the small oscillations of loading stress in long plateau stage of the tensile curve. As the strain is applied to around 0.180, Ti$_2$C sheet with full 1T' structure fails via brittle fracture to create zigzag edges, explaining the sudden drop of stress to almost zero. As a result, the applied straining energy in the highly stretched 1T' structure of Ti$_2$C sheet is pronouncedly released, followed by a rapidly reversible phase transition of to 1T' to 1T structure.

Upon zigzag directional load, however, it is observed an apparent structural transformation via locally non-uniform elongation of Ti-C bonds in the Ti$_2$C sheet as it is nonlinearly elastically strained, as illustrated by the insets of Figure 4b. Such localized non-uniform elongation of Ti-C bonds results in distortion of the hexagonal rings in Ti$_2$C sheets. Note that, in sharp contrast to the rapid structural transformation for the case under armchair directional load, this structural transformation tardily occurs, corresponding to the relative smooth tensile curve. As a result of non-uniform distribution of shear strain, highly stretched Ti$_2$C fails via ductile fracture, as shown in snapshots of Figure 4b at strains of 0.273 and 0.274. Similarly, as the fracture occurs, non-fractured region of Ti$_2$C sheet fully recovers to its origin 1T phase, as seen in the snapshot at strain of 0.274 in Figure 4b. As for fully-functionalized Ti$_2$C sheet subjected to armchair directional elastic load, it is quite uniformly stretched over the whole sheet, differing from the case of functionalization-free Ti$_2$C sheet. As the fully-functionalized Ti$_2$C sheet is over elongated, failure initially locally occurs, followed by rapid failure along the zigzag direction, as illustrated by the snapshots at strains of 0.108 and 0.109 in Figure 4c. With regard to fully-functionalized Ti$_2$C sheet subjected to zigzag directional elastic straining, it is observed structural transformation through locally non-uniform elongation of Ti-C bonds in the sheet as the applied strain reaches to critical value of around 0.110, as indicated by snapshots at strains of 0.110 and 0.120 of Figure 4d. Such structural transformation is responsible for the short plateau phase in the tensile curve. As the fully-functionalized Ti$_2$C sheet is highly zigzag directional elongation, it fails via ductile fractures, as illustrated by snapshot at strain of 0.143 in Figure 4d. Such ductile failure explains the long strain regime in the final deformational stage.

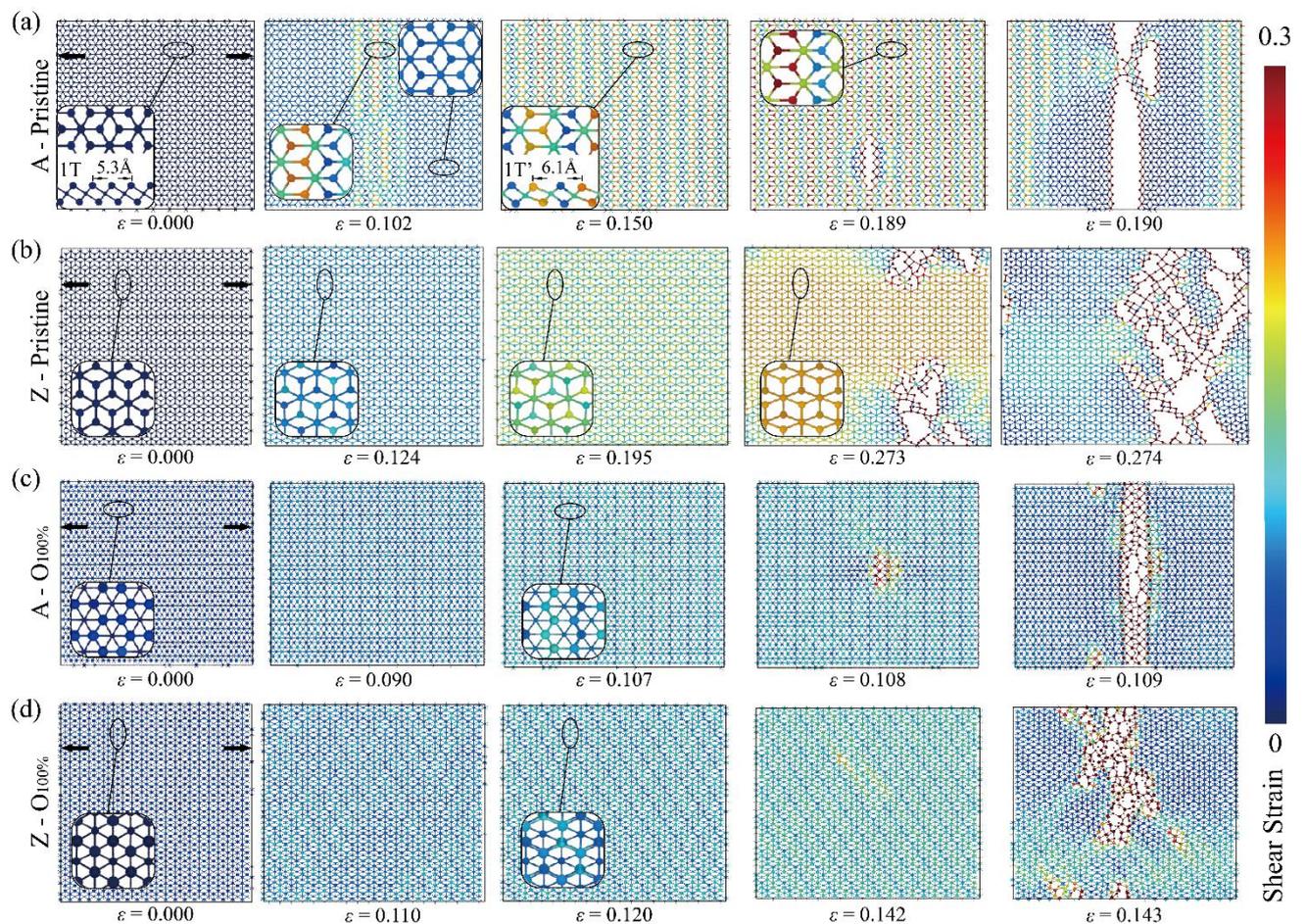

Figure 4 Representative structural development of Ti$_2$CO$_x$ systems under mechanical loads. Top-

viewed snapshots of (a) $Ti_2C$ and (b) $Ti_2CO_2$ sheets subjected to armchair directional straining. Top-viewed snapshots of (c) $Ti_2C$ and (d) $Ti_2CO_2$ sheets under zigzag directional load. The insets highlight the localized atomic structural development. The atoms are colored based on the values of shear strains.

### 3.4 Critical Bond Configurations in Functionalization-free $Ti_2C$ sheet

Upon uniaxial mechanical load, $Ti_2CO_x$ sheets are responses primarily via localized bond and bond angular deformation. Quantitatively monitoring the variations in the length and angle of bond configurations are of importance to understand the deformation mechanisms. In this work, the representative bonds and bond angles in both functionalization-free and fully-functionalized $Ti_2C$ sheets are marked (Figures 5a and b). Under armchair and zigzag directional straining, the unique tensile responses of functionalization-free $Ti_2C$ sheet are primarily attributed to the occurrence of structural transformations as illustrated by the insets in Figures 5c and f. It is revealed that as-selected six representative Ti-C bonds undergo distinct responses to the load, depending on the bond orientation to the loading direction and their location.

Figure 5d shows variations in the length of $L_a$, $L_a'$, $L_b$ and $L_b'$ bonds with strain as the functionalization-free $Ti_2C$ sheet is stretched along armchair direction. Obviously, $L_a$ and $L_a'$ bonds that make zero planar angle to the loading direction are more pronounced, indicating that they are the most load bearing bonds. As the elastic strain is applied from 0.0 to 0.10, both $L_a$ and $L_a'$ bonds are monotonically elongated from around 2.10 to 2.20 Å by 4.76%, while $L_b$ and $L_b'$ bonds that make 60° planar angle to the stretching direction are monotonically strained from around 2.10 to 2.15 Å by 2.38%. As the strain is in the range of around 0.103 to 0.140 that corresponds to phase transition of 1T to 1T' structure, there are significant changes in the length of those bonds. Interestingly, the lengths of $L_a$ and $L_b$ bonds are sharply increased by around 7.72% and 1.86%, respectively. The similar $L_a$' and $L_b$' bonds, however, reduce from 2.20-2.13 Å and 2.15-2.13 Å, respectively. Such different scenarios in the length changes of similar bonds mainly comes from the occurrence of 1T to 1T' phase transition. With further increasing strain up to fracture, both $L_a$ and $L_b$ bonds are continuously elongated, whereas for similar $L_a$' and $L_b$' bonds, there is negligible change in their bond lengths. Prior to catastrophic failure, it is identified the critical length of $L_a$ bond of about 2.48 Å.

Upon zigzag directional load, variations in the length of $L_a$, $L_b$ and $L_b$' bonds with strain are captured in Figure 5g. As is seen, $L_b$ and $L_b$' bonds that make 30° planar angle become the most load bearing bonds, while $L_a$ bond that makes zero planar angle becomes much less sensitive to the load. Prior to strain-hardening behavior (< strain of around 0.120), both $L_b$ and $L_b$' bonds show overlapped linear bond length-strain curves, namely, they are almost linearly elongated with increasing strain. Once the strain is imposed from around 0.120-0.225 that corresponds to the strain-hardening regime, it is observed separation behavior in the length of $L_b$ and $L_b$' bonds - strain curves. Such different changes in the lengths of $L_b$ and $L_b$' bonds stems from non-uniform deformation over the sheet caused by structural transformation, as illustrated by the insets of Figure 5f. With further augment of strain, both length of $L_b$ and $L_b$' bonds - strain curves are again overlapped, indicating the accomplishment of structural transformation. Prior to catastrophic failure, both $L_b$ and $L_b$' bonds show maximum bond length of around 2.33 Å that is much lower than that of $L_a$ bond under armchair directional load. However, $L_a$ bond is only elongated as the sheet is linearly elastically strained, followed by extreme long plateau phase in its bond length - strain curve.

Upon both directional loads, the imposed strain exceeds the elongation of bonds prior to fractures. This clearly indicates that stretching of bond angles also plays an important role in the in-plane deformation of functionalization-free Ti$_2$C sheet. Therefore, Figures 5e and h plot variations in the four selected bond angles with strain along armchair and zigzag directional loads, respectively. Similar to the case of bond length, the changes in bond angles strongly depend on the orientation and location of bond angles, as well as the loading direction. Upon armchair directional straining, $\beta$ and $\beta$' bond angles are identically monotonically increased with increase of elastic strain, however, there is opposite change in the $\alpha$ and $\alpha$' bond angles, differing from the case of bond length. Similar to the case of bond length, as a result of occurrence of phase transition, there is separation in the bond angle - strain curves of similar $\alpha$ and $\alpha$' ($\beta$ and $\beta$'). Prior to the catastrophic failure, $\alpha$' and $\beta$ bond angles show minimum and maximum values of approximately 88° and 107°, respectively.

Subjected to zigzag directional load, it is identified from Figure 5h that both $\beta$ and $\beta$' bond angles negligibly vary with strain from 0.00 to 0.120. Likewise, both $\beta$ and $\beta$' bond angles experience changes within strain regime of 0.120-0.225. The $\beta$ bond angle initially enlarges with increasing strain, but then decreases with further increasing strain. With regard to the $\beta$' bond angle, however, it is observed an initial reduction but then a rise within the strain regime. Such different changes in similar $\beta$ and $\beta$' bond angles again explain the structural transformation. As the sheet is further stretched, both $\beta$ and $\beta$' bond angles become identical, followed by a sudden drop as a consequence of occurrence of fractures. However, the selected $\alpha$ bond angle is monotonically increased from 93°-113° by 21.5% as the strain is increased from 0.0-0.273, followed by a sudden deep drop to original 93° due to catastrophic failure, in contrast to the case of bond length. This indicates that bond angular deformation plays more critical role in the global deformation of functionalization-free Ti$_2$C sheet under zigzag directional load.

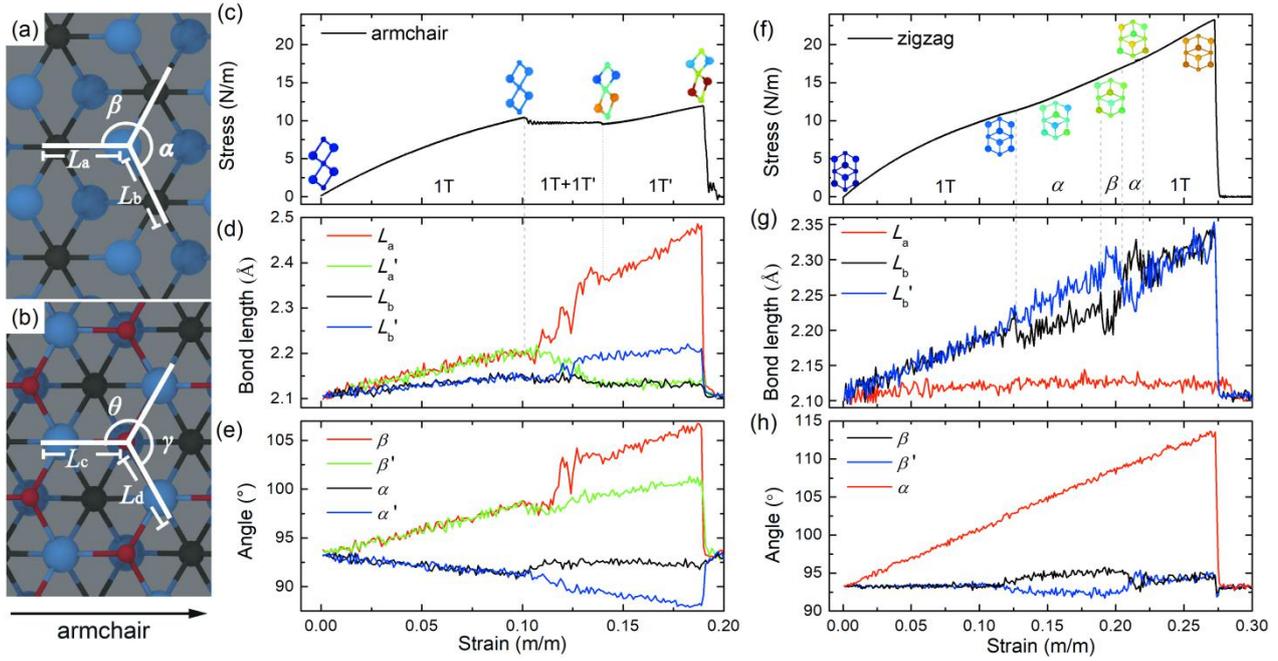

Figure 5 Kinetics in bond configurations of functionalization-free Ti$_2$C sheet. (a) and (b) Zoomed-in top-views of pristine and fully O-functionalized Ti$_2$C sheets, in which representative bonds and representative bond angles are marked for quantitatively monitoring their changes with uniaxial straining, respectively. Variations in the (c) tensile stress, (d) representative bond lengths and (e) representative bond angles of functionalization-free Ti$_2$C sheet with strain along the armchair direction. Variations in the (f) tensile stress, (g) representative bond lengths and (h) representative bond angles of functionalization-free Ti$_2$C sheet with strain along the zigzag direction, respectively. The atoms in insets are colored based on the values of shear strains.

**3.5 Critical Bond Configurations in Fully-O-functionalized Ti$_2$C sheet**

Figure 6 shows the representative bond lengths and bond angles of fully O-functionalized Ti$_2$C sheet as a function of applied strain, as well as the tensile loading curves with insets of localized atomic structures at critical strains. Note that $L_a$, $L_a$', $L_b$, $L_b$' and $\alpha$, $\alpha$', $\alpha$'', $\beta$, $\beta$', $\beta$'' are the representative Ti-C bonds and C-Ti-C bond angles, respectively, while $L_c$, $L_c$', $L_d$, $L_d$' and $\gamma$, $\gamma$', $\theta$, $\theta$' denote the representative Ti-O bonds and Ti-O-Ti bond angles, respectively. As a result of O-chemically bonded to Ti atoms on the planar surfaces, there are significant changes in the lengths of $L_a$, $L_a$', $L_b$ and $L_b$' bonds and the bond angles of $\alpha$, $\alpha$', $\alpha$'', $\beta$, $\beta$' and $\beta$'' in relaxed fully O-functionalized Ti$_2$C sheet. $L_a$, and $L_b$ bonds are increased from around 2.10 to 2.17 Å by 3.33%, while $L_a$' and $L_b$' bonds are enlarged from around 2.10 to 2.41 Å by 14.8%, indicating O-functionalization induced elongation of Ti-C bonds. With regard to the $\alpha$, $\alpha$', $\alpha$'', $\beta$ and $\beta$' bond angles, however, they are declined from around 93° to 84°, 78°, 88°, 84° and 83° by 9.67%, 16.1%, 5.38%, 9.67% and 10.8%, respectively, indicating O-functionalization induced contraction of Ti-C-Ti bond angles. Newly introduced Ti-O ($L_c$, $L_c$', $L_d$ and $L_d$') bond lengths are identical (around 1.90 Å), while as-formed $\gamma$, $\gamma$', $\theta$ and $\theta$' Ti-O-Ti bond angles are found to be approximately 105°, 105°, 107° and 107°, respectively. Despite their different bond configurations, relaxed fully-O-functionalized Ti$_2$C sheet is able to maintain 1T phase crystalline structure similar to that of functionalization-free Ti$_2$C sheet.

Upon armchair directional elastic elongation, as shown in Figures 6b and c, those representative bonds and bond angles are monotonically elongated or contracted with increasing strain, depending on their orientations and locations arranged in the sheet. Among them, it is identified that $L_c$ bond, $\theta$ and $\beta$ bond angles are the most deformational bond configurations as a result of their unique orientations and locations, suggesting that they are the main load-bearing bond configurations. Furthermore, it is observed that the maximum deformation of critical Ti-O bond (10.53 %) is nearly 2-folds of that of critical Ti-O-Ti bond angle (6.54 %), indicating that Ti-O bond plays more dominative role in the deformation of fully-O-functionalized Ti$_2$C sheet under armchair directional load, which is opposite to the case of functionalization-free Ti$_2$C sheet. By comparing with the case of functionalization-free Ti$_2$C sheet, Ti-C bonds in fully-O-functionalized Ti$_2$C sheet show much less deformational ability. The strong Ti-O bonds on the surfaces are responsible for the mechanical robustness in the armchair direction of fully-O-functionalized Ti$_2$C sheet.

With regard to zigzag directional case, as shown in Figures 6e and f, those representative bonds and bond angles show distinct deformational responses, relying on the orientation, location and type of bond configurations. For example, bonds with small orientational angles to the loading directions are elongated, while those with large orientational angles are contracted, as seen in Figure 6e. Among all bonds, both $L_d$ and $L_d$' bonds exhibit the largest elongation ability, indicating they are the major load-bearing bonds. Interestingly, $L_c$, $L_c$', $L_d$ and $L_d$' bonds show nearly monotonic changes in their bond lengths during the whole elastic load, however, there exists an obvious crossover in the bond lengths of $L_a$ and $L_b$ at critical strain of around 0.107 at which the structural transformation initiates. This clearly reveals that Ti-O bonds play negligible role in the strain-induced structural transformation of fully-O-functionalized Ti$_2$C sheet. Similarly, depending on the orientation and location, bond angles are either stretched or contracted. Unlike the case of Ti-O bonds, there are sudden changes as the structural transformation takes place, suggesting that all bond angles play a positive role in the occurrence of structural transformation. By comparison, the maximum strain of Ti-O-Ti bond angles (around 9.52 %) is significant higher than that of Ti-O bonds (about 5.26 %), indicating that specific Ti-O-Ti bond angles play a more significant role in the uniaxial deformation of fully-O-functionalized Ti$_2$C sheet.

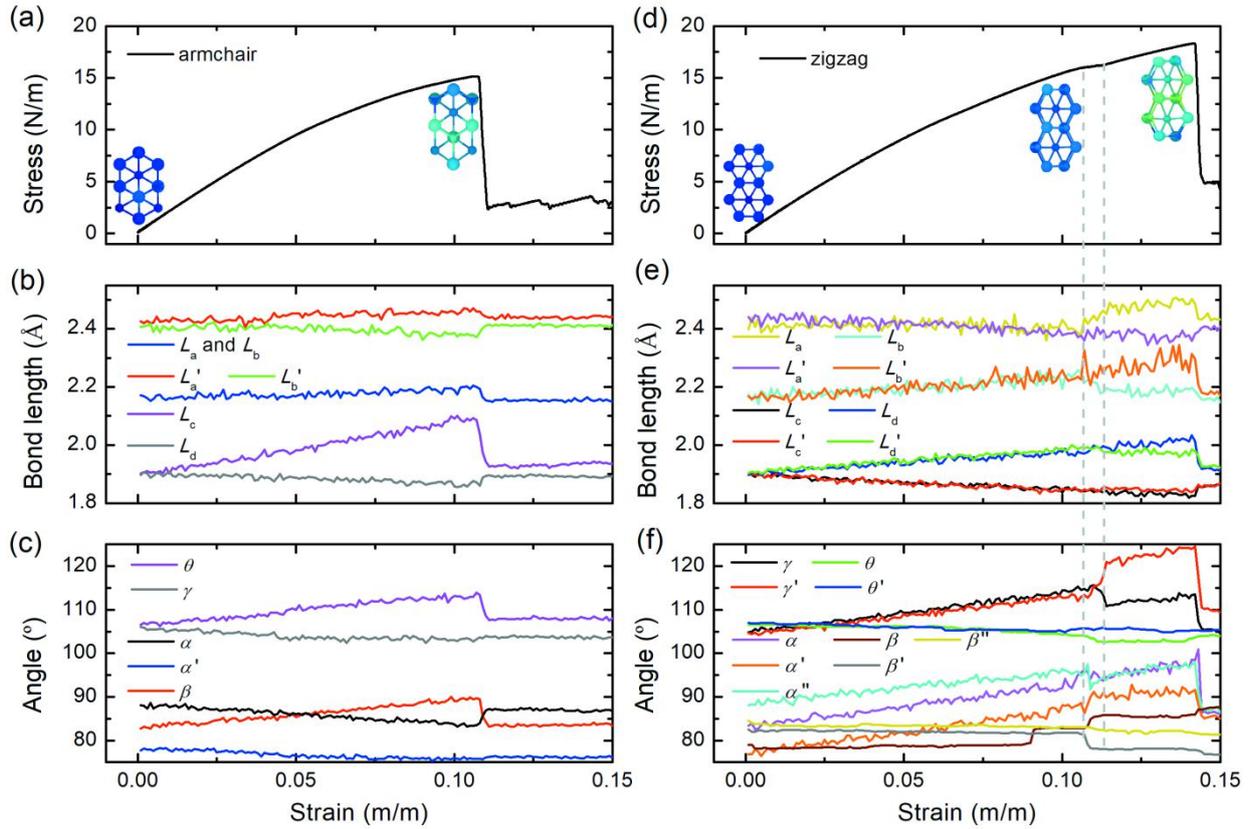

Figure 6 Kinetics in bond configurations of $Ti_2CO_2$ sheet. Variations in the (a) tensile stress, (b) representative bond lengths and (c) representative bond angles with strain along the armchair direction, respectively. Variations in the (d) tensile stress, (e) representative bond lengths and (f) representative bond angles with strain along the zigzag direction, respectively. Note that the representative bonds and representative bond angles are highlighted in Figures 5a and b.

## 4. Conclusions

In summary, MD simulations with a first-principle based ReaxFF forcefield are performed to investigate the tensile mechanical characteristics of thinnest $Ti_2C$ MXene sheet with O-functionalization randomly distributed on both planar surface. MD results show that $Ti_2CO_x$ ($x$ varies from 0 to 2) sheets exhibit unique tensile mechanical responses that can be divided into three or four deformational stages, depending on the content of O-functionalization and elongation direction. Both the first and second deformational stages are characterized by initial linear and nonlinear elastic responses, whereas for the third and fourth deformational stages, they are primarily characterized by either strain-hardening behavior or nonlinear responses with singularities (such as sudden drops of loading stress) in the curves. Surprisingly, it is identified an obvious crossover in the fundamental tensile properties of O-functionalized $Ti_2C$ MXene sheets; The Young's modulus, failure strength and failure strain are initially reduced but then are enhanced as the content of O-functionalization varies from 0.0-1.0. Upon armchair directional stretching, it is for the first time observed a phase transition from 1T to 1T' crystalline structure that has been reported in other 2D TMDs, explaining the unique nonlinearity with singularities in the tensile stress-strain curves. Subjected to zigzag directional elongation, however, it is identified two different structural transformations occurred in functionalization-free and O-functionalized $Ti_2C$ areas, respectively. The three structural transformations are able to be reversible as the straining is removed. Development of representative bonds and bond angles reveals that local transformations of specific bond and bond angular configurations are primarily responsible for those intriguing structural transformations.

**Acknowledgements**

This work is financially supported by the National Natural Science Foundation of China (Grant Nos. 11772278, 11904300 and 11502221), the Jiangxi Provincial Outstanding Young Talents Program (Grant No. 20192BCBL23029), the Fundamental Research Funds for the Central Universities (Xiamen University: Grant Nos. 20720180014, 20720180018 and 20720160088). Y. Yu and Z. Xu from Information and Network Center of Xiamen University for the help with the high-performance computer.

**References**

1.    Naguib, M., et al., *Two-dimensional nanocrystals produced by exfoliation of Ti3 AlC2*. Adv

Mater, 2011. **23**(37): p. 4248-53.

2. Naguib, M., et al., *25th anniversary article: MXenes: a new family of two-dimensional materials*. Adv Mater, 2014. **26**(7): p. 992-1005.

3. Zhu, J., et al., *Recent advance in MXenes: A promising 2D material for catalysis, sensor and chemical adsorption*. Coordination Chemistry Reviews, 2017. **352**: p. 306-327.

4. Kim, S.J., et al., *Metallic Ti3C2Tx MXene Gas Sensors with Ultrahigh Signal-to-Noise Ratio*. ACS Nano, 2018. **12**(2): p. 986-993.

5. Wu, M., et al., *Ti3C2 MXene-Based Sensors with High Selectivity for NH3 Detection at Room Temperature*. ACS Sensors, 2019. **4**(10): p. 2763-2770.

6. Lukatskaya, M.R., et al., *Cation Intercalation and High Volumetric Capacitance of Two-Dimensional Titanium Carbide*. Science, 2013. **341**(6153): p. 1502.

7. Fan, G., et al., *Magnetic, recyclable PtyCo1−y/Ti3C2X2 (X = O, F) catalyst: a facile synthesis and enhanced catalytic activity for hydrogen generation from the hydrolysis of ammonia borane*. New Journal of Chemistry, 2017. **41**(7): p. 2793-2799.

8. Han, M., et al., *Beyond Ti3C2Tx: MXenes for Electromagnetic Interference Shielding*. ACS Nano, 2020. **14**(4): p. 5008-5016.

9. Wang, X., et al., *Atomic-Scale Recognition of Surface Structure and Intercalation Mechanism of Ti3C2X*. Journal of the American Chemical Society, 2015. **137**(7): p. 2715-2721.

10. Han, F., et al., *Boosting the Yield of MXene 2D Sheets via a Facile Hydrothermal-Assisted Intercalation*. ACS Applied Materials & Interfaces, 2019. **11**(8): p. 8443-8452.

11. Okubo, M., et al., *MXene as a Charge Storage Host*. Accounts of Chemical Research, 2018. **51**(3): p. 591-599.

12. Xie, X., et al., *Porous Ti3C2Tx MXene for Ultrahigh-Rate Sodium-Ion Storage with Long Cycle Life.* ACS Applied Nano Materials, 2018. **1**(2): p. 505-511.

13. Wan, Q., S. Li, and J.-B. Liu, *First-Principle Study of Li-Ion Storage of Functionalized Ti2C Monolayer with Vacancies.* ACS Applied Materials & Interfaces, 2018. **10**(7): p. 6369-6377.

14. Dall'Agnese, Y., et al., *Two-Dimensional Vanadium Carbide (MXene) as Positive Electrode for Sodium-Ion Capacitors.* The Journal of Physical Chemistry Letters, 2015. **6**(12): p. 2305-2309.

15. Lipatov, A., et al., *Elastic properties of 2D $Ti_3C_2T_x$ MXene monolayers and bilayers.* 2018. **4**(6): p. eaat0491.

16. Guo, Z., et al., *Flexible two-dimensional Tin+1Cn (n = 1, 2 and 3) and their functionalized MXenes predicted by density functional theories.* Physical Chemistry Chemical Physics, 2015. **17**(23): p. 15348-15354.

17. Hu, T., et al., *Quantifying the rigidity of 2D carbides (MXenes).* Physical Chemistry Chemical Physics, 2020. **22**(4): p. 2115-2121.

18. Khaledialidusti, R., B. Anasori, and A. Barnoush, *Temperature-dependent mechanical properties of Tin+1CnO2 (n = 1, 2) MXene monolayers: a first-principles study.* Physical Chemistry Chemical Physics, 2020. **22**(6): p. 3414-3424.

19. Bai, Y., et al., *Dependence of elastic and optical properties on surface terminated groups in two-dimensional MXene monolayers: a first-principles study.* RSC Advances, 2016. **6**(42): p. 35731-35739.

20. Plummer, G., et al., *Nanoindentation of monolayer Ti C T MXenes via atomistic simulations:*

*The role of composition and defects on strength.* Computational Materials Science, 2019. **157**: p. 168-174.

21. Borysiuk, V.N., V.N. Mochalin, and Y. Gogotsi, *Molecular dynamic study of the mechanical properties of two-dimensional titanium carbides Ti(n+1)C(n) (MXenes).* Nanotechnology, 2015. **26**(26): p. 265705.

22. Borysiuk, V.N., V.N. Mochalin, and Y. Gogotsi, *Bending rigidity of two-dimensional titanium carbide (MXene) nanoribbons: A molecular dynamics study.* Computational Materials Science, 2018. **143**: p. 418-424.

23. Xie, Y., et al., *Role of surface structure on Li-ion energy storage capacity of two-dimensional transition-metal carbides.* J Am Chem Soc, 2014. **136**(17): p. 6385-94.

24. Khazaei, M., et al., *Novel Electronic and Magnetic Properties of Two-Dimensional Transition Metal Carbides and Nitrides.* Advanced Functional Materials, 2013. **23**(17): p. 2185-2192.

25. Ashton, M., et al., *Predicted Surface Composition and Thermodynamic Stability of MXenes in Solution.* The Journal of Physical Chemistry C, 2016. **120**(6): p. 3550-3556.

26. Hu, T., et al., *Chemical Origin of Termination-Functionalized MXenes: Ti3C2T2 as a Case Study.* The Journal of Physical Chemistry C, 2017. **121**(35): p. 19254-19261.

27. Li, J., et al., *Thermal stability of two-dimensional Ti2C nanosheets.* Ceramics International, 2015. **41**(2, Part A): p. 2631-2635.

28. Enyashin, A.N. and A.L. Ivanovskii, *Structural and Electronic Properties and Stability of MXenes Ti2C and Ti3C2 Functionalized by Methoxy Groups.* The Journal of Physical Chemistry C, 2013. **117**(26): p. 13637-13643.

29. Zhao, H., et al., *Anomalous thermal stability in supergiant onion-like carbon fullerenes.*

Carbon, 2018. **138**: p. 243-256.

30. Sui, C., et al., *Morphology-Controlled Tensile Mechanical Characteristics in Graphene Allotropes.* ACS Omega, 2017. **2**(7): p. 3977-3988.

31. Feng, C., et al., *Morphology- and dehydrogenation-controlled mechanical properties in diamond nanothreads.* Carbon, 2017. **124**: p. 9-22.

32. Fu, Y., et al., *The effects of morphology and temperature on the tensile characteristics of carbon nitride nanothreads.* Nanoscale, 2020. **12**(23): p. 12462-12475.

33. Osti, N.C., et al., *Effect of Metal Ion Intercalation on the Structure of MXene and Water Dynamics on its Internal Surfaces.* ACS Applied Materials & Interfaces, 2016. **8**(14): p. 8859-8863.

34. Zhang, D., et al., *Computational Study of Low Interlayer Friction in Ti$_{n+1}$C$_n$ (n = 1, 2, and 3) MXene.* ACS Applied Materials & Interfaces, 2017. **9**(39): p. 34467-34479.

35. Wei, C. and C. Wu, *Nonlinear fracture of two-dimensional transition metal carbides (MXenes).* Engineering Fracture Mechanics, 2020. **230**: p. 106978.

36. Song, S., et al., *Room Temperature Semiconductor–Metal Transition of MoTe2 Thin Films Engineered by Strain.* Nano Letters, 2016. **16**(1): p. 188-193.

37. Lin, Y.-C., et al., *Atomic mechanism of the semiconducting-to-metallic phase transition in single-layered MoS2.* Nature Nanotechnology, 2014. **9**(5): p. 391-396.

38. Duerloo, K.-A.N., Y. Li, and E.J. Reed, *Structural phase transitions in two-dimensional Mo- and W-dichalcogenide monolayers.* Nature Communications, 2014. **5**(1): p. 4214.

39. Zhao, Y., et al., *Strain-Controllable Phase and Magnetism Transitions in Re-Doped MoTe2 Monolayer.* The Journal of Physical Chemistry C, 2020. **124**(7): p. 4299-4307.